\documentclass[aps,pre,twocolumn,showpacs,superscriptaddress]{revtex4}
\usepackage{graphics,amsmath,amssymb}
\usepackage{dcolumn}
\usepackage{bm}
\newcommand{\sinp}{\affiliation{Saha Institute of Nuclear Physics,  Block-AF, Sector-I Bidhannagar, Kolkata-700064, India.}}
\newcommand{\snb}{\affiliation{Satyendra Nath Bose National Centre for Basic Sciences Block-JD, Sector-III, Salt Lake, Kolkata-700098, India.}}

\begin{document}
\title{Pareto Law in a Kinetic Model of Market with Random Saving Propensity}

\author{Arnab Chatterjee}%
\email{arnab@cmp.saha.ernet.in}
\sinp
\author{Bikas K. Chakrabarti}%
\sinp
\author{S. S. Manna}%
\snb

\begin{abstract}
We have numerically simulated the ideal-gas models of trading markets, where
each agent is identified with a gas molecule and each trading as an elastic or
money-conserving two-body collision. Unlike in the ideal gas, we introduce
(quenched) saving propensity of the agents, distributed widely between the
agents ($0 \le \lambda < 1$). The system remarkably self-organizes to a
critical Pareto distribution of money $P(m) \sim m^{-(\nu + 1)}$ with
$\nu \simeq 1$. We analyse the robustness (universality) of the distribution
in the model. We also argue that although the fractional saving ingredient
is a bit unnatural one in the context of gas models, our model is the simplest
so far, showing self-organized criticality, and combines two century-old
distributions: Gibbs (1901) and Pareto (1897) distributions.
\end{abstract}

\maketitle

\noindent
Considerable investigations have already been made to study the nature
of income or wealth distributions in various economic communities,
in particular, in different countries. For more than a hundred years,
it is known that the probability distribution $P(m)$ for income or
wealth of the individuals in the market decreases with the wealth
$m$ following a power law, known as Pareto law \cite{pareto}:
\begin{equation}
\label{par}
P(m)\propto m^{-(1+\nu) },
\end{equation}

\noindent where the value of the exponent \( \nu  \) is found to
lie between 1 and 2 \cite{olids,fuji,ls}. It is also known that typically
less than 10\% of the population in any country possesses about 40\%
of the wealth and follow the above power law. The rest of the low-income
group population, in fact the majority, clearly follows a different
law, identified very recently to be the Gibbs distribution 
\cite{marjit,dy2,anbkcepjb}. Studies on real data show that the high 
income group indeed follow Pareto law, with $\nu$ varying from $1.6$
for USA \cite{dy2}, to $1.8-2.2$ in Japan \cite{fuji}. The value of $\nu$
thus seem to vary a little from economy to economy.

We have studied here numerically a gas model of a trading market. We have 
considered the effect of saving propensity of the traders. The saving 
propensity is assumed to have a randomness. Our observations indicate that 
Gibbs and Pareto distributions fall in the same category and can appear 
naturally in the century-old and well-established kinetic theory of gas 
\cite{land}: Gibbs distribution for no saving and Pareto distribution for 
agents with quenched random saving propensity. Our model study also indicates 
the appearance of self-organized criticality \cite{bak} in the simplest model 
so far, namely in the kinetic theory of gas models, when the stability effect 
of savings \cite{sam} is incorporated.

We consider an ideal-gas model of a closed economic system 
where total money $M$ and total number of agents $N$ is fixed. No production 
or migration occurs and the only economic activity is confined to trading. 
Each agent $i$, individual or corporate, possess money $m_i(t)$ at time $t$.
In any trading, a pair of traders $i$ and $j$ randomly exchange their money 
\cite{marjit,anbkcepjb,dy}, such that their total money is (locally) conserved 
and none end up with negative money ($m_i(t) \ge 0$, i.e, debt not allowed):
\begin{equation}
\label{consv}
m_i(t) + m_j(t) = m_i(t+1) + m_j(t+1);
\end{equation}

\noindent time ($t$) changes by one unit after each trading.
The steady-state ($t \rightarrow \infty$) distribution of money is Gibbs one: 
\begin{equation}
\label{gibbs}
P(m)=(1/T)\exp(-m/T);T=M/N. 
\end{equation}

Hence, no matter how uniform or justified the initial distribution is, the
eventual steady state corresponds to Gibbs distribution where most of the
people have got very little money. 
This follows from the conservation of money and additivity of entropy:
\begin{equation}
\label{prob}
P(m_1)P(m_2)=P(m_1+m_2).
\end{equation}

\noindent
This steady state result is quite robust and realistic too!
In fact, several variations of the trading, and of the `lattice' 
(on which the agents can be put and each agent trade with its 
`lattice neighbors' only), whether compact, fractal or small-world like
\cite{olids}, leaves the distribution unchanged. Some other variations 
like random sharing of an amount $2m_2$ only (not of $m_1 + m_2$)
when $m_1 > m_2$ (trading at the level of lower economic class in the trade),
lead to even drastic situation: all the money in the market drifts to one 
agent and the rest become truely pauper \cite{anijmpc,hayes}.

In any trading, savings come naturally \cite{sam}.
A saving propensity factor $\lambda$ is therefore introduced in the same model 
\cite{anbkcepjb} (see \cite{dy} for model without savings), where each trader 
at time $t$ saves a fraction $\lambda$ of its money $m_i(t)$ and trades 
randomly with the rest: 
\begin{equation}
\label{delm}
m_{i}(t+1)=m_{i}(t)+\Delta m;\quad m_{j}(t+1)=m_{j}(t)-\Delta m 
\end{equation}
\noindent where
\begin{equation}
\label{eps}
\Delta m=(1-\lambda )[\epsilon \{m_{i}(t)+m_{j}(t)\}-m_{i}(t)],
\end{equation}

\noindent
$\epsilon$ being a random fraction, coming from the stochastic nature
of the trading.

The market (non-interacting at $\lambda =0$ and $1$) becomes `interacting' 
for any non-vanishing $\lambda (<1)$: For fixed $\lambda$ (same for all 
agents), the steady state distribution $P_f(m)$ of money is exponentially 
decaying on both sides with the most-probable money per agent shifting away 
from $m=0$ (for $\lambda =0$) to $M/N$ as $\lambda \rightarrow 1$ 
(Fig. 1(a)). This self-organizing feature of the market,
induced by sheer self-interest of saving by each agent without any global 
perspective, is quite significant as the fraction of paupers decrease with 
saving fraction $\lambda$ and most people end up with some fraction of the 
average money in the market (for $\lambda \rightarrow 1$, the socialists' 
dream is achieved with just people's self-interest of saving!).
Interestingly, self-organisation also occurs in such market models when there
is restriction in the commodity market \cite{acspbkc}.
Although this fixed saving propensity does not give yet the Pareto-like
power-law distribution, the Markovian nature of the scattering or trading
processes (eqn. (\ref{prob})) is lost and the system becomes co-operative.
Indirectly through \(\lambda\), the agents get to know (start interacting with)
each other and the system co-operatively self-organises towards a most-probable
distribution (\(m_p \ne 0\)).

\vskip 0.30in
{\centering \resizebox*{8cm}{4cm}{\includegraphics{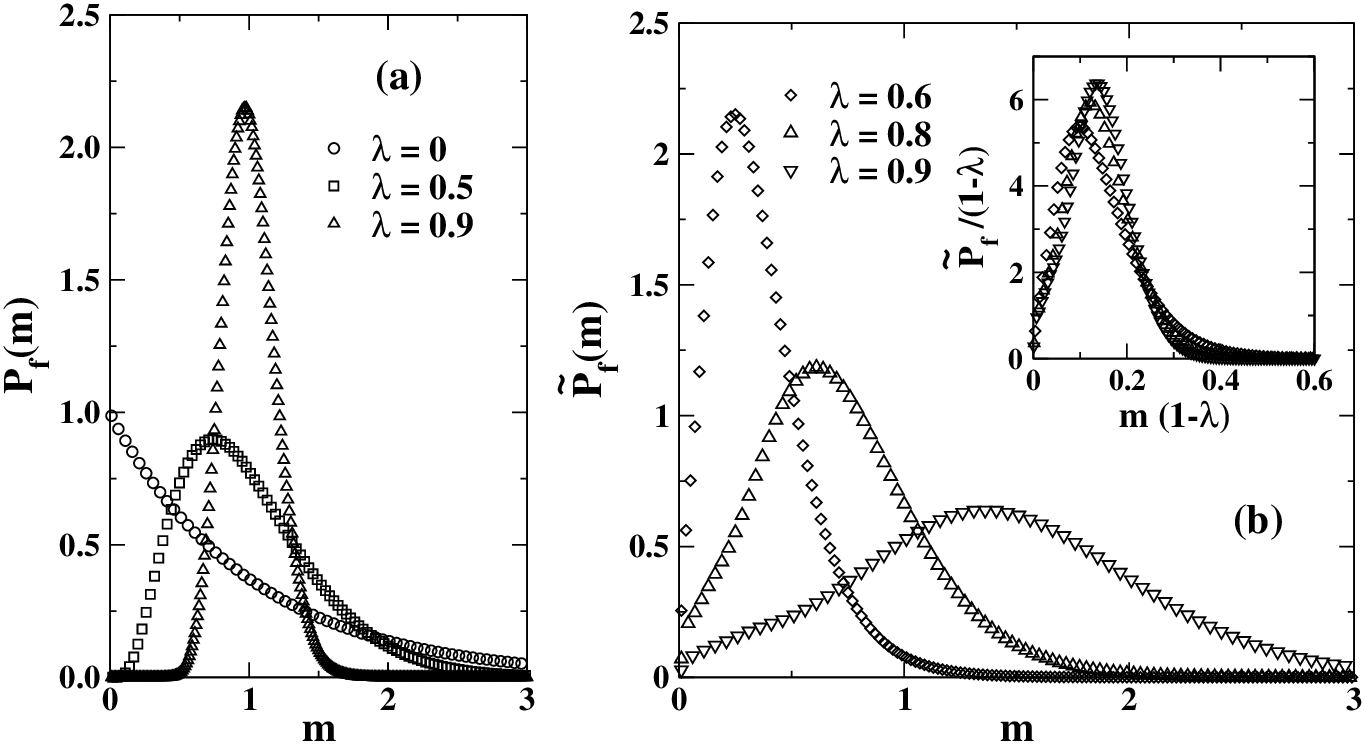}} \par}
\vskip 0.2in
\noindent {\footnotesize FIG.1: 
Steady state money distribution (a) $P_f(m)$ for the fixed (same for all 
agents) $\lambda$ model, and (b) $\tilde{P_f}(m)$ for some typical values of
$\lambda$ in the distributed $\lambda$ model. The data is collected from the 
ensembles with $N=200$ agents. The inset in (b) shows the scaling behavior of
$\tilde{P_f}(m)$. For all cases, agents 
start with average money per agent $M/N = 1$.
}{\footnotesize \par}
\vskip 0.1in

In a real society or economy, $\lambda$ is a very inhomogeneous parameter:
the interest of saving varies from person to person.
We move a step closer to the real situation where saving factor $\lambda$ is 
widely distributed within the population. One again follows the same trading
rules as before, except that
\begin{equation}
\label{lrand}
\Delta m=\epsilon(1-\lambda_{j})m_{j}(t)-(1-\lambda _{i})(1 - \epsilon)m_{i}(t)
\end{equation}

\noindent 
here; $\lambda _{i}$ and $\lambda _{j}$ being the saving
propensities of agents $i$ and $j$. The agents have fixed (over time) saving
propensities, distributed independently, randomly and uniformly (white)
within an interval $0$ to $1$ (see \cite{bkcsv} for preliminary results): 
agent $i$ saves a random fraction 
$\lambda_i$ ($0 \le \lambda_i < 1$) and this $\lambda_i$ value is quenched 
for each agent ($\lambda_i$ are independent of trading or $t$).
Starting with an arbitrary initial (uniform or random) distribution of 
money among the agents, the market evolves with the tradings. At each time, 
two agents are randomly selected and the money exchange among them occurs, 
following the above mentioned scheme. We check for the steady state, by 
looking at the stability of the money distribution in successive 
Monte Carlo steps $t$. Eventually, after a typical relaxation time 
($\sim 10^5$ for $N=200$ and uniformly distributed $\lambda$) dependent
on $N$ and the distribution of $\lambda$, the money distribution becomes
stationary. After this, we average the money distribution over $\sim 10^3$ 
time steps. Finally we take configurational average over $\sim 10^5$ 
realizations of the $\lambda$ distribution to get the money distribution
$P(m)$. It is found to follow a strict power-law decay. 
This decay fits to Pareto 
law (\ref{par}) with $\nu = 1.02 \pm 0.02$ (Fig. 2). Note, for 
finite size $N$ of the market, the distribution has a narrow initial growth 
upto a most-probable value $m_p$ after which it falls off with a power-law 
tail for several decades. As can be seen from the inset of Fig. 2, this 
Pareto law (with $\nu \simeq 1$) covers the entire range in $m$ of the 
distribution $P(m)$ in the limit $N \rightarrow \infty$. We checked that 
this power law is extremely robust: apart from the uniform $\lambda$ 
distribution used in the simulations in Fig. 2, we also checked the results 
for a distribution 
\begin{equation}
\label{lam0}
\rho(\lambda) \sim |\lambda_0-\lambda|^\alpha,\quad \lambda_0 \ne 1, \quad 0<\lambda<1,
\end{equation}

\noindent
of quenched $\lambda$ values among the agents. The Pareto law with $\nu=1$ is
universal for all $\alpha$. The data in Fig. 2 corresponds to 
$\lambda_0 = 0$, $\alpha = 0$. For negative $\alpha$ values, however, 
we get an initial (small $m$) Gibbs-like decay in $P(m)$ (see Fig. 3).

\vskip 0.3in
{\centering \resizebox*{8cm}{5.5cm}{\includegraphics{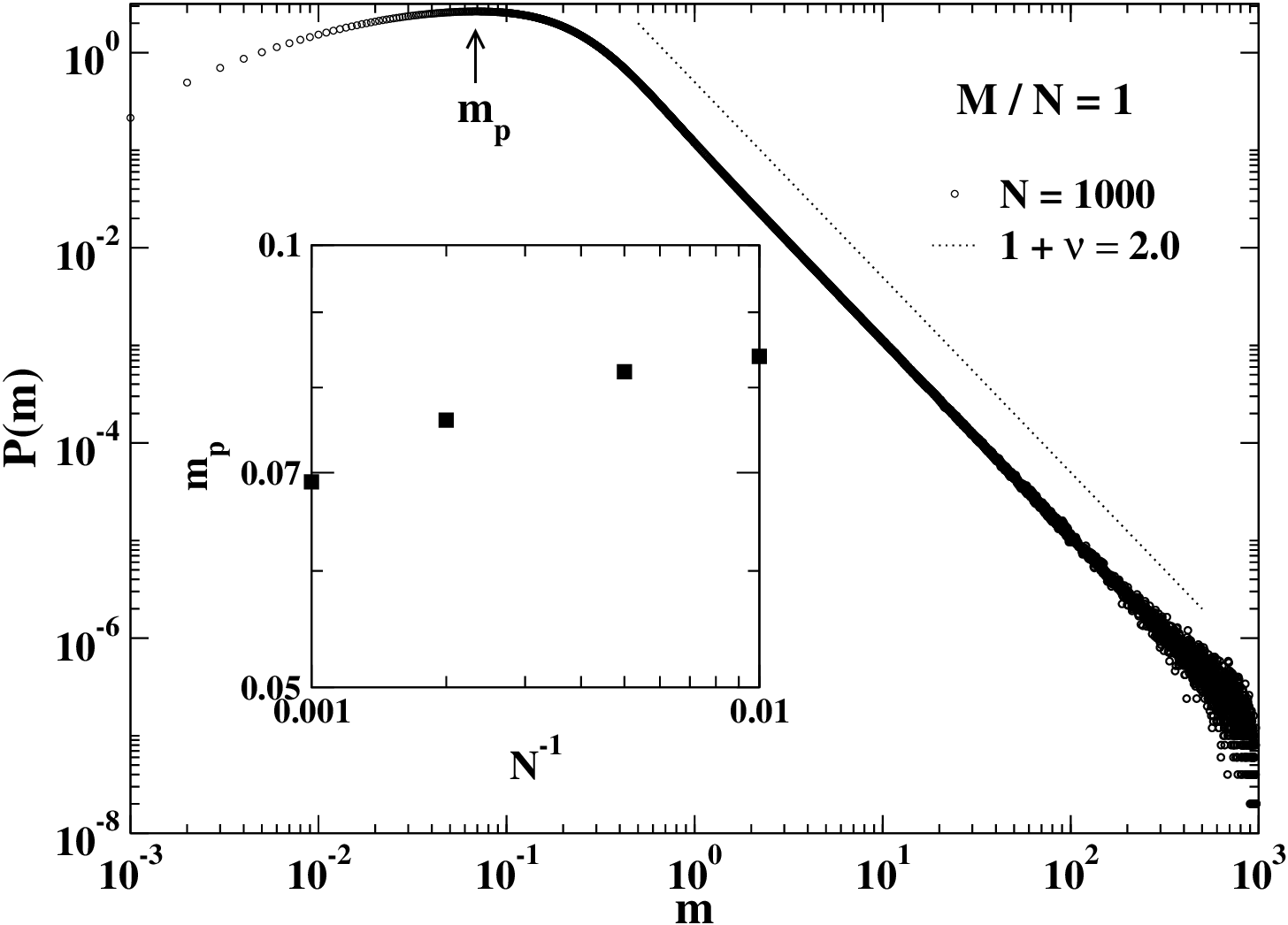}} \par}
\vskip 0.2in
\noindent {\footnotesize FIG.2: 
Steady state money distribution $P(m)$ in the model for distributed $\lambda$
($0 \le \lambda < 1$) for $N = 1000$ agents. Inset shows that the most 
probable peak $m_p$ shifts towards $0$ (indicating the same power law for the 
entire range of $m$) as $N \rightarrow \infty$: results for four typical system 
sizes $N = 100, 200, 500, 1000$ are shown.
For all cases, agents play with average money per agent $M/N = 1$.
}{\footnotesize \par}
\vskip 0.05in

\vskip 0.2in
{\centering \resizebox*{8cm}{6.0cm}{\includegraphics{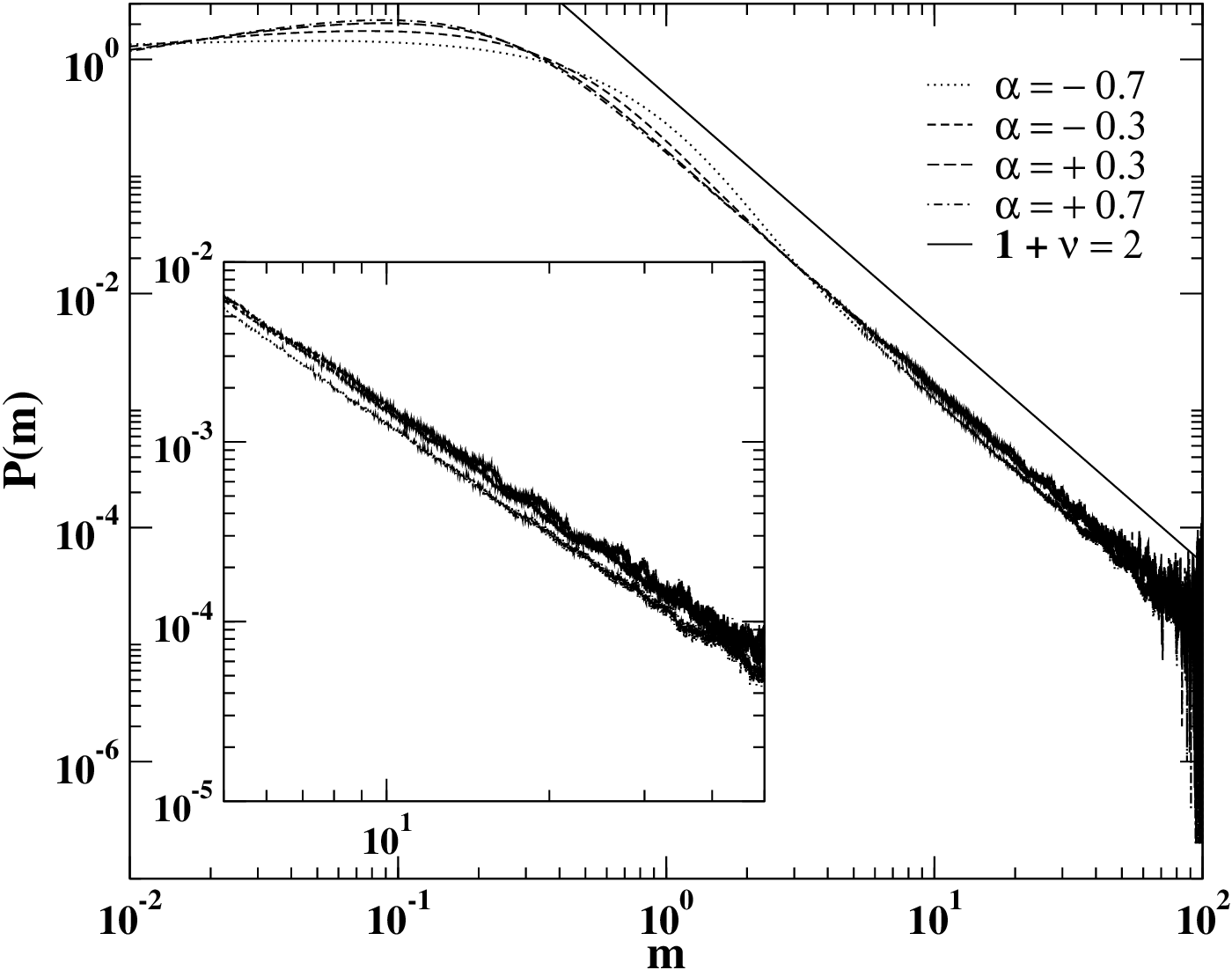}} \par}
\vskip 0.2in
\noindent {\footnotesize FIG.3: 
Steady state money distribution $P(m)$ in the model for $N = 100$ agents with
$\lambda$ distributed as  $\rho (\lambda) \sim \lambda^\alpha$ with different
values of $\alpha$. The inset shows the region of validity of Pareto law
with $1+\nu=2$.
For all cases, agents play with average money per agent $M/N = 1$.
}{\footnotesize \par}
\vskip 0.05in

In case of uniform distribution of saving propensity $\lambda$ 
($0 \le \lambda <1$), the individual money distribution $\tilde{P_f}(m)$ for 
agents with any particular $\lambda$ value, although differs considerably, 
remains non-monotonic: similar to that for fixed $\lambda$ market 
with $m_p(\lambda)$ shifting with $\lambda$ (see Fig. 1). Few subtle points may
be noted though: while for fixed $\lambda$ the $m_p(\lambda)$ were all 
less than of the order of unity (Fig. 1(a)), for distributed $\lambda$ case
$m_p(\lambda)$ can be considerably larger and can approach to the order of $N$
for large $\lambda$ (see Fig. 1(b)). The other important difference is in the
scaling behavior of $\tilde{P_f}(m)$, as shown in the inset of Fig. 1(b).
In the distributed $\lambda$ ensemble, $\tilde{P_f}(m)$ appears to have a very
simple scaling:
\begin{equation}
\label{scale}
\tilde{P}_f(m) \sim (1-\lambda) \mathcal F (m(1-\lambda)),
\end{equation}

\noindent
for $\lambda \rightarrow 1$, where the scaling function ${\cal F} (x)$ has 
non-monotonic variation in $x$. The fixed (same for all agents) $\lambda$ 
income distribution $P_f(m)$ do not have any such comparative scaling property. 
It may be noted that a small difference exists between the
ensembles considered in Fig 1(a) and 1(b): while $\int mP_f(m)dm=M$ 
(independent of $\lambda$), $\int m \tilde{P_f}(m)dm$ is not a constant and
infact approaches to order of $M$ as $\lambda \rightarrow 1$.
There is also a marked qualitative difference in fluctuations (see Fig. 4):
while for fixed $\lambda$, the fluctuations in time (around the most-probable 
value) in the individuals' money $m_i(t)$ gradually decreases with increasing 
$\lambda$, for quenched distribution of $\lambda$, the trend gets reversed 
(see Fig. 4).

\vskip 0.4in
{\centering \resizebox*{8cm}{6.8cm}{\includegraphics{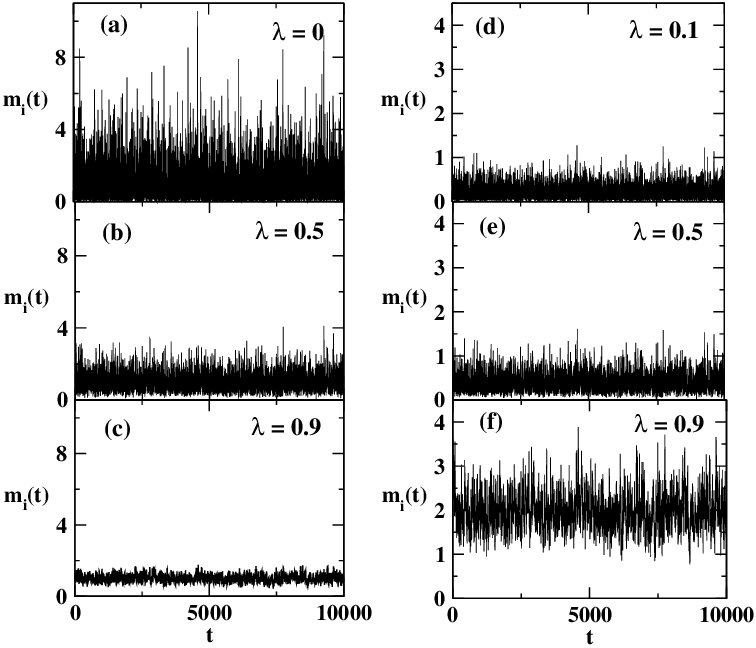}} \par}
\vskip 0.1in
\noindent {\footnotesize FIG.4: 
Money of the $i$-th trader with time. For fixed $\lambda$ for all agents in the
market: (a) with $\lambda = 0$, (b) $\lambda = 0.5$, (c) $\lambda = 0.9$. 
For distributed $\lambda$ ($0 \le \lambda < 1$): for agents with 
(d) $\lambda = 0.1$, (e) $\lambda = 0.5$ and (f) $\lambda = 0.9$ in the market. 
All data are for $N=200$ ($M/N=1$).
}{\footnotesize \par}
\vskip 0.1in

\vskip 0.3in
{\centering \resizebox*{8cm}{8cm}{\includegraphics{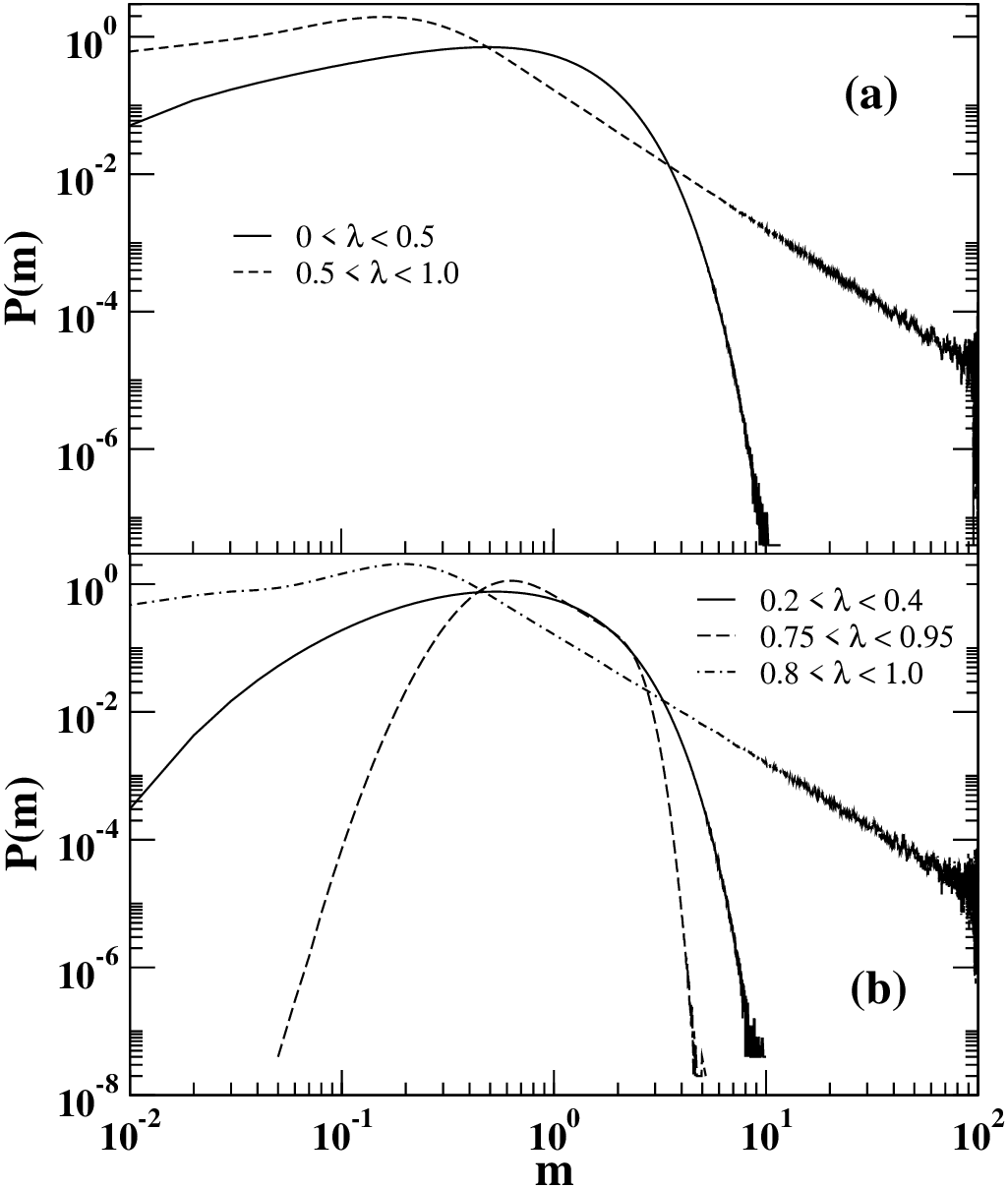}} \par}
\vskip 0.1in
\noindent {\footnotesize FIG.5: 
Money distribution in cases where saving propensity $\lambda$ is distributed 
uniformly within a range of values: (a) When width of $\lambda$ distribution 
is $0.5$, money distribution shows power-law for $0.5 < \lambda < 1.0$; 
(b) When width of $\lambda$ distribution is $0.2$, money distribution starts 
showing power-law when $0.7 < \lambda < 0.9$. Note, the exponent 
$\nu \simeq 1$ in all these cases when the power law is seen.
All data are for $N=100$ ($M/N=1$).
}{\footnotesize \par}
\vskip 0.1in

We now investigate on the range of distribution of the saving propensities in
a certain interval $a<\lambda_i<b$, where, $0<a<b<1$. For uniform distribution
within the range, we observe the
appearance of the same power law in the distribution but for a narrower
region. As may be seen from Fig. 5, as $a \rightarrow b$, the power-law
behavior is seen for values $a$ or $b$ approaching more and more towards unity:
For the same width of the interval $|b-a|$, one gets power-law (with
same $\nu$) when $b \rightarrow 1$. This indicates, for fixed $\lambda$, 
$\lambda=0$ corresponds to Gibbs distribution, and one gets Pareto law when 
$\lambda$ has got non-zero width of its distribution extending upto 
$\lambda = 1$. This of course indicates a crucial
role of these high saving propensity agents: the power law behavior is truely 
valid upto the asymptotic limit if $\lambda = 1$ is included. Indeed, had we 
assumed $\lambda_0=1$ in (\ref{lam0}), 
the Pareto exponent $\nu$ immediately switches over to $\nu=1+\alpha$. 
Of course, $\lambda_0 \ne 1$ in (\ref{lam0}) leads to the universality of the 
Pareto distribution with $\nu = 1$ (independent of $\lambda_0$ and $\alpha$). 
Indeed this can be easily rationalised from the scaling behavior (\ref{scale}):
$P(m) \sim \int_0^1 \tilde{P_f}(m)\rho(\lambda)d\lambda$ $\sim$ $m^{-2}$ for
$\rho(\lambda)$ given by (\ref{lam0}) and $m^{-(2+\alpha)}$ if 
$\lambda_0=1$ in (\ref{lam0}) (for large $m$ values).

\vskip 0.2in
{\centering \resizebox*{8cm}{8cm}{\includegraphics{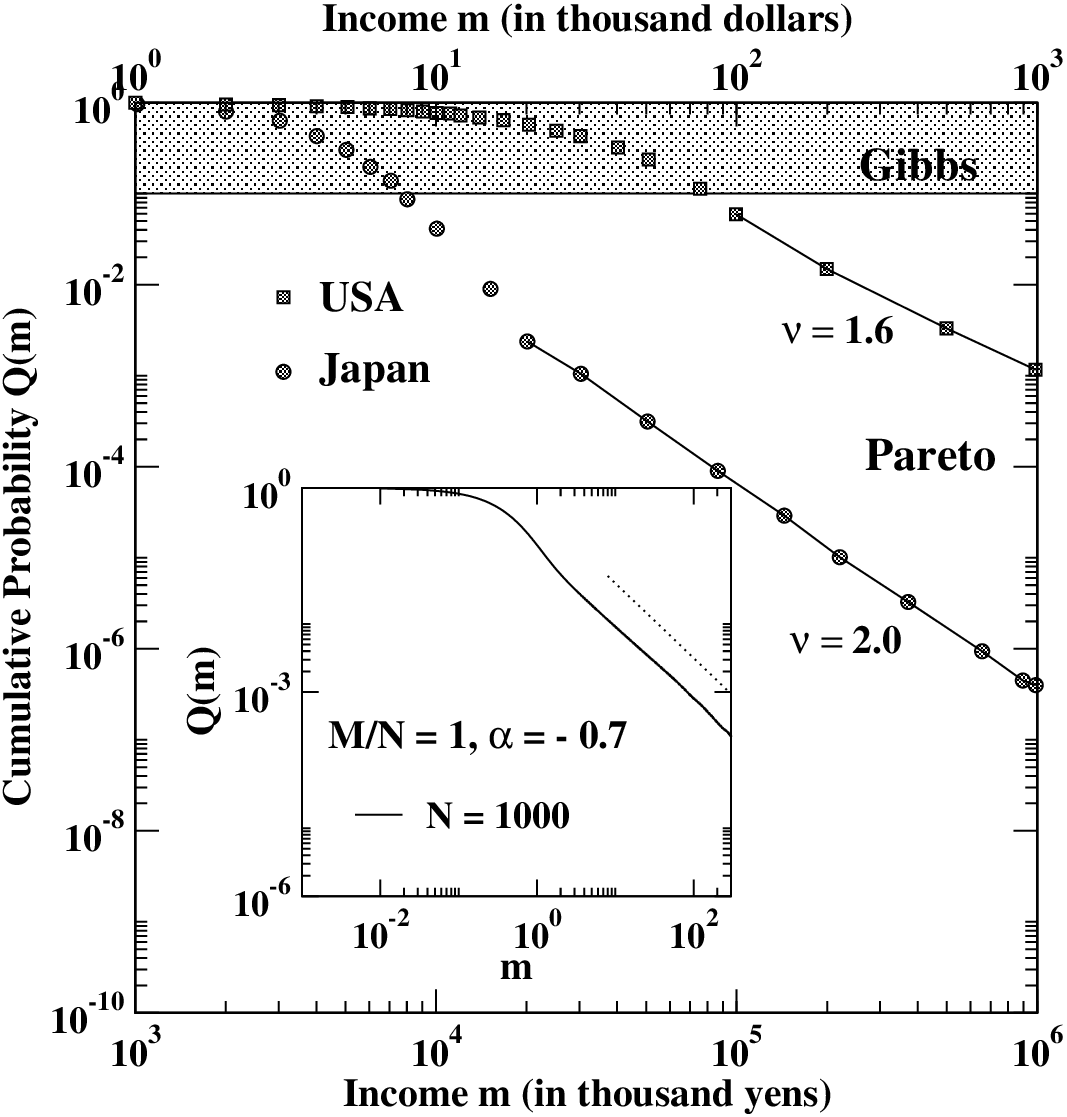}} \par}
\vskip 0.2in
\noindent {\footnotesize FIG.6: 
Cumulative distribution $Q(m)=\int_m^\infty P(m)dm$ of wealth $m$ in USA 
\cite{dy2} in 1997 and Japan \cite{fuji} in 2000. Low-income group follow Gibbs 
law (shaded region) and the rest (about 5\%) of the rich population follow 
Pareto law. The inset shows the cumulative distribution for a model market 
where the saving propensity of the agents is distributed following (\ref{lam0})
with $\lambda_0=0$ and $\alpha=-0.7$.
The dotted line (for large $m$ values) corresponds to $\nu = 1.0$.
}{\footnotesize \par}

These model income distributions $P(m)$ compare
very well with the wealth distributions of various countries: Data suggests
Gibbs like distribution in the low-income range \cite{dy2} (more than 90\% of
the population) and Pareto-like in the high-income range \cite{fuji}
(less than 10\% of the population) of various countries (Fig. 6).
In fact, we have compared one model simulation of the market with saving 
propensity of the agents distributed following (\ref{lam0}), with $\lambda_0=0$
and $\alpha=-0.7$. This model result is shown in the inset of Fig. 6.
The qualitative resemblance of the model income distribution with the real
data for Japan and USA in recent years is quite intriguing. In fact, for
negative $\alpha$ values in (\ref{lam0}), the density of traders with low
saving propensity is higher and since $\lambda=0$ ensemble yields 
Gibbs-like income distribution (\ref{gibbs}), we see an initial Gibbs-like
distribution which crosses over to Pareto distribution (\ref{par}) with 
$\nu=1.0$
for large $m$ values. The position of the crossover point depends on the
magnitude of $\alpha$. The important point to note is that any distribution
of $\lambda$ near $\lambda=1$, of finite width, eventually gives Pareto law
for large $m$ limit. The same kind of crossover behavior (from Gibbs to Pareto)
can also be reproduced in a model market of mixed agents where $\lambda=0$
for a finite fraction of population and $\lambda$ is distributed uniformly over
a finite range near $\lambda=1$ for the rest of the population.

We also considered annealed randomness in the saving propensity $\lambda$:
here $\lambda_i$ for any agent $i$ changes from one value to another within 
the range $0 \le \lambda_i < 1$, after each trading. Numerical studies for 
this annealed model did not show any power law behavior for $P(m)$; rather 
it again becomes exponentially decaying on both sides of a most-probable value. 

We have numerically simulated here ideal-gas like models of trading markets, 
where each agent is identified with a gas molecule and each trading as an 
elastic or money-conserving two-body collision. Unlike in the ideal gas, 
we introduce (quenched) saving propensity of the agents, distributed widely 
between the agents ($0 \le \lambda < 1$). For quenched random variation of 
$\lambda$ among the agents the system remarkably self-organizes to a 
critical Pareto distribution (\ref{par}) of money with $\nu \simeq 1.0$ 
(Fig. 2). The exponent is quite robust: for savings distribution
$\rho(\lambda) \sim |\lambda_0-\lambda|^\alpha$, 
$\lambda_0 \ne 1$, one gets the same Pareto law with $\nu = 1$ (independent
of $\lambda_0$ or $\alpha$). It may be noted that the trading market model
we have talked about here has got some apparent limitations. The stochastic 
nature of trading assumed here in the trading market, through the random 
fraction $\epsilon$ in (\ref{eps}), is of course not very straightforward 
as agents apparently go for trading with some definite purpose (utility 
maximization of both money and commodity). We are however, looking only at 
the money transactions between the traders. In this sense, the income 
distribution we study here essentially corresponds to `paper money', and not 
the `real wealth'. However, even taking money and commodity together, one can 
argue (see \cite{hayes}) for the same stochastic nature of the tradings, due to 
the absence of `just pricing' and the effects of bargains in the market.

Apart from the intriguing observation that Gibbs (1901) and Pareto (1897) 
distributions fall in the same category and can appear naturally in the 
century-old and well-established kinetic theory of gas, that this model 
study indicates the appearance of self-organized criticality in 
the simplest (gas) model so far, when the stability effect of savings 
incorporated, is remarkable.


\newpage
\end{document}